\documentstyle[11pt,moriond]{article}
\input psfig.tex

\def\Journal#1#2#3#4{{#1} {\bf #2}, #3 (#4)}

\def\AAP{\em Astron. Astrophys.}
\def\AAS{\em Astron. Astrophys. Suppl. Ser.}
\def\AJ{\em Astron. J.}
\def\APP{\em Astropart. Phys.} 
\def\ANNREV{\em Ann. Rev. Astron. Astrophys.}
\def\APJ{\em Astrophys. J.}
\def\APJS{\em Astrophys. J. Suppl.}

\def\MN{\em Mon. Not. R. Astr. Soc.}
\def\PASP{\em Publ. Astr. Soc. Pacific.}
\def\NAT{\em Nature}
\def\SAIT{\em Mem. Soc. Astron. It.} 
\def\ltsima{$\; \buildrel < \over \sim \;$}
\def\simlt{\lower.5ex\hbox{\ltsima}}            
\def\gtsima{$\; \buildrel > \over \sim \;$}
\def\simgt{\lower.5ex\hbox{\gtsima}}            

\def\be{\begin{equation}}
\def\ee{\end{equation}}
\def\bea{\begin{eqnarray}}
\def\eea{\end{eqnarray}}
\begin{document}
\vspace*{4cm}
\title{GAMMA-RAY EMITTING AGN AND UNIFIED SCHEMES~\footnote{Invited Review, to 
appear in {\it Very High Energy Phenomena in the Universe}, XXXIInd Moriond 
Conference}}

\author{PAOLO PADOVANI}

\address{Dipartimento di Fisica, II Universit\`a di Roma ``Tor Vergata''\\
Via della Ricerca Scientifica 1, I-00133 Roma, Italy\\
WWW: http://itovf2.roma2.infn.it/padovani/padovani.html}

\maketitle\abstracts{Active Galactic Nuclei (AGN) are now known to be strong
$\gamma$-ray emitters. After briefly describing the different classes of AGN 
and the basic tenets of unified schemes, I discuss the role of blazars (that is
BL Lacs and flat-spectrum radio quasars) as $\gamma$-ray sources. The main
properties of blazars and their connection with relativistic beaming are then
summarized. Finally, I address the question of why blazars, despite being
extreme and very rare objects, are the only AGN detected at very high ($E > 100$
MeV) energies, and touch upon the relevance of TeV astronomy for AGN research.} 

\section{Active Galactic Nuclei}\label{sec:agn}

Active Galactic Nuclei (AGN) are extragalactic sources, in some cases clearly
associated with nuclei of galaxies (although generally the host galaxy seems
to be too faint to be seen), whose emission is dominated by non-stellar
processes in some waveband(s) (typically but not exclusively the optical). One
important feature of AGN is the fact that their emission covers the whole
electromagnetic spectrum, from the radio to the $\gamma$-ray band, sometimes
over almost 20 orders of magnitude in frequency. 

It is now well established that AGN are strong $\gamma$-ray ($E > 100$ MeV)
emitters. To be more specific: 1. at least 40\% of all EGRET sources are
AGN~\cite{k97} (some more AGN are certainly present amongst the still
unidentified sources) and these make up almost 100\% of all extragalactic
sources (the only exceptions being the Large Magellanic Cloud and possibly
Centaurus A); 2. {\it all} detected AGN are blazars, that is BL Lacertae
objects (BL Lacs) or flat-spectrum radio quasars (FSRQ). \footnote{The term
``blazar'' is here given a wider meaning than the one sometimes implied, which
is restricted to highly polarized quasars (HPQ) and/or optically violently
variable (OVV) quasars. The reason is that there is increasing evidence that
these categories and the flat-spectrum radio quasars, which reflect different
empirical definitions, refer to more or less the same class of sources. That
is, the majority of flat-spectrum radio quasars tend to show rapid variability
and high polarization.} To appreciate the relevance of the latter point, we
will first have to tackle the subject of AGN classification. 

\subsection{AGN Classification}\label{sec:class}

The large number of classes and subclasses appearing in AGN literature could
disorientate physicists or even astronomers working in other fields. A
simplified classification, however, can be made based on only two parameters,
that is radio-loudness and the width of the emission lines, as summarized in
Table 1 (see e.g., Urry \&  Padovani~\cite{up95}). 

\begin{table}[t]
\caption{AGN Taxonomy: A Simplified Scheme.}\label{tab:taxo}
\vspace{0.4cm}
\begin{center}
\begin{tabular}{|l|l|l|l|}         
\hline
{\bf Radio Loudness} & \multicolumn{3}{c|}{\bf Optical Emission Line
Properties} \\ \hline 

 & {\bf Type 2} (Narrow Line) & {\bf Type 1} (Broad Line) & {\bf Type 0} 
(Unusual) \\ 
 & & & \\
Radio-quiet: & Seyfert 2 & Seyfert 1 &  \\
 & & & \\
             &           & QSO       &  \\ 
 & & & \\
\hline
 & & & \\
Radio-loud: & NLRG $\cases {{\rm FR~I} \cr ~ \cr {\rm FR~II} \cr}$ & 
BLRG & Blazars $\cases {{\rm BL~Lacs} \cr ~ \cr {\rm (FSRQ)} \cr}$ \\

            &    & SSRQ & \\
            &    & FSRQ & \\
 & & & \\
\hline
 & \multicolumn{3}{c|}{decreasing angle to the line of sight $\longrightarrow$} \\
\hline
\end{tabular}
\end{center}
\end{table}

Although it was the strong radio emission of some quasars which led to their
discovery more than 30 years ago, it soon became evident that the majority of
quasars were actually radio-quiet, that is most of them were not detected by
the radio telescopes of the time. It then turned out that radio-quiet did not
mean radio-silent, that is even radio-quiet AGN can be detected in the radio
band. Why then the distinction? If one plots radio luminosity versus optical
luminosity for complete samples of optically selected sources, it looks like
there are two populations, the radio quiet one having, for the same optical
power, a radio power which is about $3 - 4$ orders of magnitudes smaller.
The distribution of the luminosity ratio $L_{\rm r}/L_{\rm opt}$ for complete
samples, including the upper limits on the radio luminosity, appears to be
bimodal, with a dividing line at a value $L_{\rm r}/L_{\rm opt} \sim 10$
(e.g., Stocke et al.;~\cite{st92} both luminosities are in units of power/Hz).
It would therefore be better to call the two classes ``radio-strong'' and
``radio-weak'' but the original names are still used. Note that only about $10
- 15\%$ of AGN are radio-loud. 

The other main feature used in AGN classification is the width, in case they 
are present, or the absence, of emission lines. These are produced by the
recombination of ions of various elements (most notably H, He, C, N, O, Ne,
Mg, Fe). Their width is due to the Doppler effect thought to result from
more or less ordered motion around the central object. AGN are then divided in
Type 1 (broad-lined) and Type 2 (narrow-lined) objects according to their
line-widths, with 1000 km/s (full width half maximum) being the dividing
value. Some objects also exist with unusual emission line properties, such as
BL Lacs, which have very weak emission lines with typical equivalent widths (a
measure of the ratio between line and continuum luminosity) $< 5$ \AA. 

As illustrated in Table 1, we then have radio-quiet Type 2 and Type 1 AGN,
that is Seyfert 2 galaxies and Seyfert 1 galaxies/radio-quiet quasars (QSO)
respectively. Radio-loud Type 2 AGN are radio galaxies (sometimes also called
narrow-line radio galaxies [NLRG] to distinguish them from the broad-lined
ones), classified as Fanaroff-Riley~\cite{fr74} I and II (FR I and II)
according to their radio morphology (connected with their radio power), while 
radio-loud Type 1 AGN are broad-line radio galaxies (BLRG) and radio quasars.
Finally, radio-loud sources with very weak emission lines are known as BL
Lacertae objects, from the name of the class prototype, which was originally
presumed to be a variable star in the Lacerta constellation. 

Concentrating on the radio-loud sources, to which most of this paper is
devoted, the BLRG are, at least in my view, simply local versions of radio
quasars where we can detect the host galaxy, as Seyfert 1 galaxies are local
versions of QSO (the reasons why we do not see the high-redshift counterparts 
of Seyfert 2 galaxies are discussed by Padovani~\cite{pad97}). Radio quasars
are generally divided into steep-spectrum radio quasars (SSRQ) and
flat-spectrum radio quasars (FSRQ), according to the value of their radio
spectral index at a few GHz ($\alpha_{\rm r} = 0.5$ being usually taken as the
dividing line, with $f_{\nu} \propto \nu^{-\alpha}$). This distinction
reflects the size of the radio emitting region. In fact, radio emission in
these sources is explained in terms of synchrotron radiation (that is
radiation from relativistic particles moving in a magnetic field), which for
extended regions has a relatively steep spectrum ($\alpha_{\rm r} \sim 0.7$).
On the other hand, nuclear, compact emission has a flatter spectrum, thought
to be the result of the superposition of various self-absorbed components. The
flat radio spectrum then indicates that nuclear emission dominates over the
more extended emission, generally associated with the so-called ``radio-lobes.'' 
In fact, flat-spectrum quasars are generally core-dominated in the radio band,
that is emission from the core is much stronger than emission from the 
extended regions, unlike for example SSRQ or narrow-line radio galaxies which
are both lobe-dominated. Note that even though FSRQ have strong broad lines
they are also included in the ``Type 0'' column in Table 1 because their
multifrequency spectra are dominated by non-thermal emission as in BL Lac
objects. 

\subsection{Unified Schemes}\label{sec:unif}

\begin{figure}[p]
\hskip 2.0truecm 
\psfig{figure=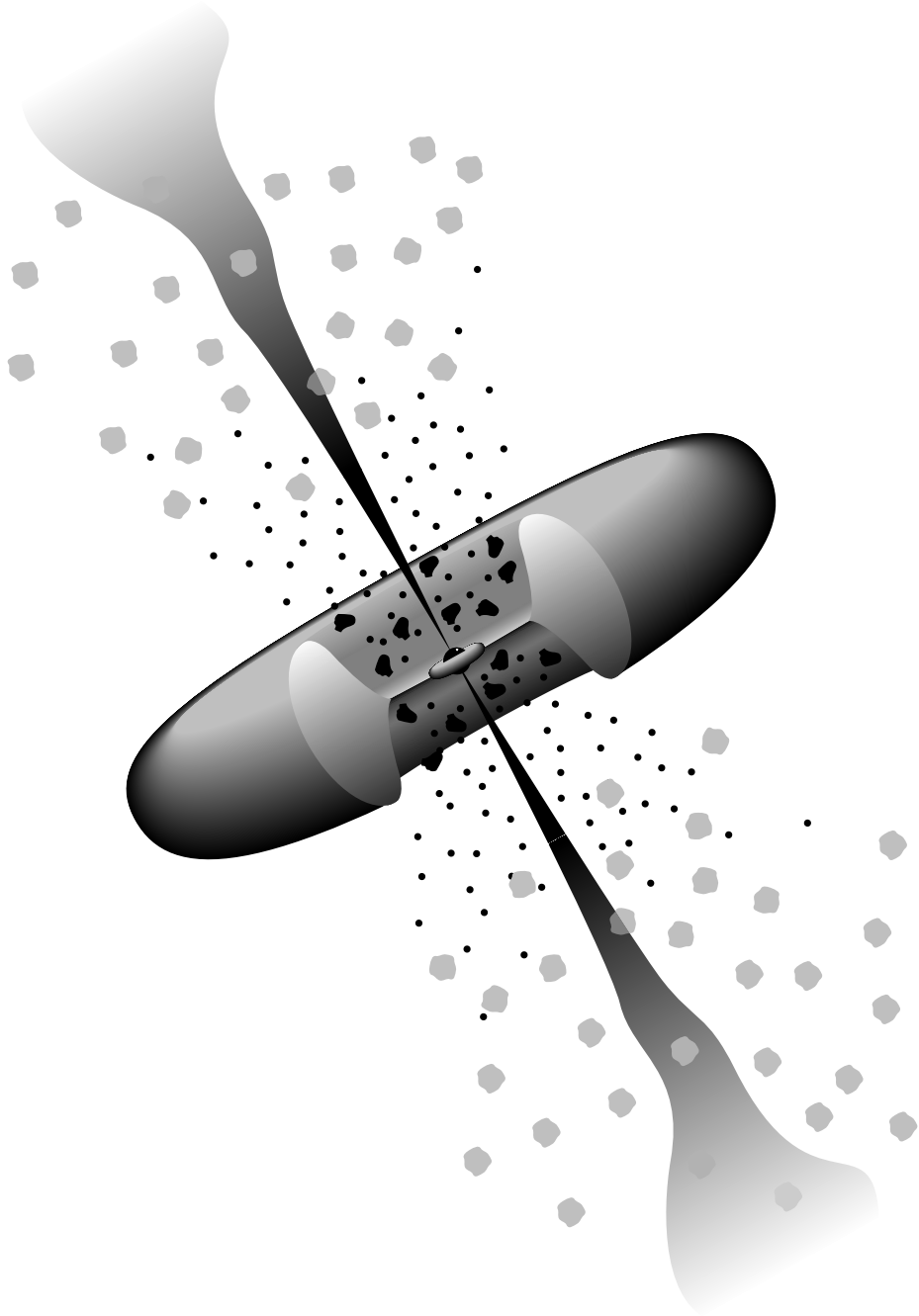,height=18truecm}
\caption{A schematic diagram of the current paradigm for radio-loud AGN (not
to scale). Surrounding the central black hole is a luminous accretion disk.
Broad emission lines are produced in clouds (dark spots) orbiting above the
disk and perhaps by the disk itself. A thick dusty torus (or warped disk)
obscures the broad-line region from transverse lines of sight; some continuum
and broad-line emission can be scattered into those lines of sight by hot
electrons (black dots) that pervade the region. A hot corona above the
accretion disk may also play a role in producing the hard X-ray continuum.
Narrow lines are produced in clouds (grey spots) much farther from the central
source. Radio jets, shown here as the diffuse jets characteristic of
low-luminosity, or FR~I-type, radio sources, emanate from the region near the
black hole, initially at relativistic speeds (Urry \& Padovani 1995; copyright 
Astronomical Society of the Pacific, reproduced with permission).} 
\label{fig:unif}
\end{figure}

All this might seem complicated, but in recent years we have developed a
consistent scenario which at least explains the Type 1/Type 2 distinction. We
have in fact come to understand that some classes of apparently different AGN
(therefore classified under different names) might actually be intrinsically 
the same class of objects seen at different angles with the line of sight (see
for example Antonucci~\cite{a93}~and Urry \& Padovani~\cite{up95} and
references therein). 

The main idea, based on various observations and summarized in 
figure~\ref{fig:unif}, is that emission in the inner parts of AGN is highly
anisotropic. The current paradigm for AGN includes a central engine, possibly
a massive black hole, surrounded by an accretion disk and by fast-moving
clouds, probably under the influence of the strong gravitational field,
emitting Doppler-broadened lines. More distant clouds emit narrower lines.
Absorbing material in some flattened configuration (usually idealized as a
toroidal shape) obscures the central parts, so that for transverse lines of
sight only the narrow-line emitting clouds are seen (Type 2 AGN), whereas the
near-infrared to soft-X-ray nuclear continuum and broad-lines are visible only
when viewed face-on (Type 1 AGN). In radio-loud objects we have the additional
presence of a relativistic jet, roughly perpendicular to the disk, which
produces strong anisotropy and amplification of the continuum emission
(``relativistic beaming''), which I will discuss in more detail later on. For
reasons still unclear, in BL Lac objects the emission lines are extremely
weak, and the continuum is very strong and non-thermal (i.e., due to
synchrotron and, at higher frequencies, inverse Compton emission or perhaps
hadronic processes). 

This axisymmetric model of AGN implies widely different observational
properties (and therefore classifications) at different aspect angles. Hence
the need for ``Unified Schemes'' which look at intrinsic, isotropic
properties, to unify fundamentally identical (but apparently different)
classes of AGN. Seyfert 2 galaxies have therefore been ``unified'' with
Seyfert 1 galaxies, whilst low-luminosity (FR I) and high-luminosity (FR II)
radio galaxies have been unified with BL Lacs and radio quasars respectively
(see Antonucci~\cite{a93} and Urry \& Padovani~\cite{up95} and references
therein). In other words, BL Lacs are thought to be FR I radio galaxies with
their jets at relatively small ($\simlt 20 - 30^{\circ}$) angles w.r.t. the
line of sight. Similarly, we believe FSRQ to be FR II radio galaxies oriented
at small ($\simlt 15^{\circ}$) angles, while SSRQ should be at angles
in between those of FSRQ and FR II's ($15 \simlt \theta \simlt 40^{\circ}$).
Blazars are then a special class of AGN which we think have their jets 
practically oriented towards the observer. 

In general, different AGN components are important at different wavelengths.
Namely: 1. the jet emits non-thermal radiation, via electromagnetic
(synchrotron and inverse Compton) and perhaps hadronic processes, all the way
from the radio to the $\gamma$-ray band;~\cite{ma97}$^{,}$~\cite{p97} 2. the
accretion disk probably emits thermal radiation, peaked in
optical/ultraviolet/soft-X-ray band; 3. the absorbing material (torus) will
emit predominantly in the infrared. These different components are apparent,
for example, in the multifrequency spectrum of 3C 273,~\cite{l95} the first
quasar to be discovered and one of the best studied. 

At this point one might ask: what has all this to do with $\gamma$-ray
emission? The answer in the next section.  

\section{AGN as $\gamma$-ray Sources: the Role of Blazars}\label{sec:role}

According to Unified Schemes, blazars are that special class of radio-loud AGN
with their jets practically pointing towards the observer, and therefore constitute
a relatively rare class of objects. Radio-loud AGN make up only $\sim 10 -
15\%$ of all AGN (e.g., Kellermann et al.~\cite{k89}), while a generous upper
limit to the fraction of blazars amongst radio sources is 50\% (as inferred,
for example, from the fraction of FSRQ and BL Lacs in the 1 Jy
catalogue~\cite{s94} which, being a high-frequency radio catalogue, is biased
towards flat-spectrum sources). It then follows that blazars make up at most
5\% of all AGN, but more likely even less than that.

Mattox et al.~\cite{m97} have identified 42 high-confidence EGRET sources 
(mainly from the Second EGRET catalogue~\cite{t95}) with AGN, all of them 
blazars. If the probability of detecting an AGN with EGRET were independent of
the class, then in this list we would expect at maximum 2 blazars, with most
sources being associated with radio-quiet AGN. Instead, we have 100\% blazars
and 0\% other sources. In particular, no radio-quiet AGN has been detected so 
far by EGRET. Blazar $\gamma$-ray luminosities are in the range $10^{45} - 
10^{49}$ erg/s and in many cases the output in $\gamma$-rays dominates the 
total (bolometric) luminosity. 

To find out what is so special about blazars we need to have a closer look at 
their properties. 

\subsection{Blazar Properties and Relativistic Beaming}\label{subsec:prop}

The main properties of blazars can be summarized as follows:

\begin{itemize}
\item radio loudness;
\item rapid variability (high $\Delta L/\Delta t$);
\item high and variable polarization ($P_{\rm opt} > 3\%$);
\item smooth, broad, non-thermal continuum; 
\item compact, flat-spectrum radio emission ($f_{\rm core} \gg f_{\rm
extended}$);
\item superluminal motion in sources with multiple-epoch Very Large Baseline
Interferometry (VLBI) maps.  
\end {itemize}

The last property might require some explanation. The term ``superluminal
motion'' describes proper motion of source structure (traditionally mapped at
radio wavelengths) that, when converted to an apparent speed $v_{\rm app}$,
gives $v_{\rm app} > c$. This phenomenon occurs for emitting regions moving at
very high (but still subluminal) speeds at small angles to the line of 
sight.~\cite{r66} Mirabel~\cite{mi97} describes examples of superluminal
motion in our own Galaxy. 

\begin{figure}
\hskip 2.0truecm 
\psfig{figure=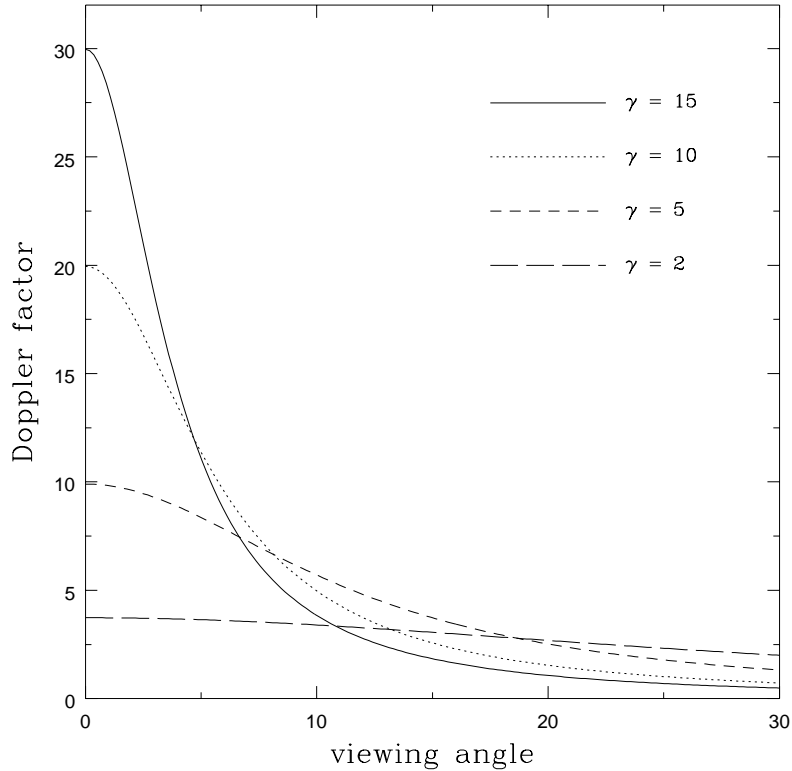,width=11truecm}
\caption{The dependence of the Doppler factor $\delta$ on the angle to the 
line of sight. Different curves correspond to different Lorentz factors: from 
the top down, $\gamma = 15$ (solid line), $\gamma = 10$ (dotted line), $\gamma 
= 5$ (short-dashed line), $\gamma = 2$ (long-dashed line).}
\label{fig:doppler}
\end{figure}

All these properties are consistent with {\it relativistic beaming}, that is
with bulk relativistic motion of the emitting plasma towards the observer.
There are by now various arguments in favour of relativistic beaming in
blazars, summarized for example by Urry \& Padovani.~\cite{up95} Beaming has
enormous effects on the observed luminosities. Adopting the usual definition
of the relativistic Doppler factor $\delta = [\gamma (1 - \beta \cos
\theta)]^{-1}$, $\gamma = (1-\beta^2)^{-1/2}$ being the Lorentz factor, with
$\beta = v/c$ and $\theta$ the angle w.r.t. the line of sight, and applying
simple relativistic transformations, it turns out that the {\it observed}
luminosity at a given frequency is related to the {\it emitted} luminosity in
the rest frame of the source via 

\begin{equation}
L_{\rm obs} = \delta^p L_{\rm em} \quad ,
\end{equation}

with $p = 2+\alpha$ or $3+\alpha$ respectively in the case of a continuous jet
or a moving sphere~\cite{up95} ($\alpha$ being the spectral index), although
other values are also possible.~\cite{lb85} For $\theta \sim 0^{\circ}$,
$\delta \sim 2 \gamma$ (figure~\ref{fig:doppler}) and the observed luminosity
can be amplified by factors of thousands (for $\gamma \sim 5$ and $p \sim 3$,
which are typical values). That is, for jets pointing almost towards us we can
overestimate the emitted luminosity typically by three orders of magnitude.
Apart from this amplification, beaming also gives rise to a strong collimation
of the radiation, which is larger for higher $\gamma$
(figure~\ref{fig:doppler}): $\delta$ decreases by a factor $\sim 2$ from its
maximum value at $\theta \sim 1/\gamma$ and consequently the inferred
luminosity goes down by $2^p$. For example, if $\gamma \sim 5$ the
luminosity of a jet pointing $\sim 11^{\circ}$ away from our line of sight is
already about an order of magnitude smaller (for $p = 3$) than that of a jet
aiming straight at us. 

All this is very relevant to the issue of $\gamma$-ray emission from blazars.
In fact, if blazars were not beamed, we would not see any $\gamma$-ray photons
from them! The qualitative explanation is relatively simple: in sources as
compact as blazars all $\gamma$-ray photons would be absorbed through
photon-photon collisions with target photons in the X-ray band. The end 
product would be electron-positron pairs. But if the radiation is beamed then
the luminosity/radius ratio, which is the relevant parameter, is smaller by a
factor $\delta^{p+1}$ and the $\gamma$-ray photons manage to escape from the
source. More formally, it can be shown~\cite{m92} that the condition that the
optical depth to photon-photon absorption $\tau_{\gamma \gamma}(x)$ is less
than 1 implies (under the assumption that X-ray emission arises from the same
region as the $\gamma$-ray emission) 

\begin{equation}
\delta > C \left({L_{48}\over \Delta t_{\rm d}}\right)^{1/
(p+1)} \left({x \over 10^4}\right)^{\alpha_{\rm x}/(p+1)} \quad ,
\end{equation}

where $L_{48} \equiv L_{\gamma}/(10^{48}$ erg/s), $\Delta t_{\rm d}$ is the
$\gamma$-ray variability time scale in days (which is used to estimate the
source size), $x \equiv h\nu/m_{\rm e} c^2$, $\alpha_{\rm x}$ is the X-ray
spectral index, and $C$ is a numerical constant $\approx 10$. In other words,
transparency for the $\gamma$-ray photons {\it requires} a relatively large
Doppler factor for most blazars~\cite{dg95} and therefore relativistic beaming.

\section{The Importance of Being a Blazar}\label{sec:impo}

We can now address the main question of this paper: why have blazars been
detected by EGRET? There are, I believe, at least three reasons, which have to
do with the fact that blazars are characterized by:

\begin{enumerate}
\item high-energy particles, which can produce GeV photons; 
\item relativistic beaming, to avoid photon-photon collision and amplify the
flux;
\item strong non-thermal (jet) component. 
\end{enumerate}

Point number one is obvious. We know that in some blazars synchrotron emission
reaches at least the optical/ultraviolet range, which reveals the presence of
high-energy electrons which can produce $\gamma$-rays via inverse Compton
emission (although hadronic processes can also be important or even
dominant~\cite{ma97}). Point number two is vital, as described in the previous
section, not only to enable the $\gamma$-ray photons to escape from the
source, but also to amplify the flux and therefore make the source more easily
detectable. Point number three is also very important. $\gamma$-ray emission
is clearly non-thermal (although we still do not know for sure which processes
are responsible for it) and therefore related to the jet component. The
stronger the jet component, the stronger the $\gamma$-ray flux. 

Having understood why blazars have been detected by EGRET, one could also ask:
why have not {\it all} blazars been detected? Many blazars with radio
properties similar to those of the detected sources, in fact, still have only
upper limits in the EGRET band. This problem has been addressed, for example,
by von Montigny et al.,~\cite{mo95} who suggest as possible solutions
variability (only objects flaring in the $\gamma$-ray band can be detected) or
a $\gamma$-ray beaming cone which either points in a different direction or is
more narrowly beamed that the radio one (see also Salamon \&
Stecker~\cite{ss94} and Dermer~\cite{d95}). These can certainly be viable
explanations, but one should also note that even a moderate dispersion in the
values of the parameters required for $\gamma$-ray emission described above
(particle energy, Doppler factor, and non-thermal component strength) can
easily imply the non-detection of some sources and the detection of others
with similar radio properties.  

Do the points discussed above also explain why EGRET has not detected any of
the more numerous radio-quiet AGN? Yes, although not all of them might be
essential in this case. As radio-emission (at least in radio-loud AGN) is
certainly non-thermal, while the optical/ultraviolet emission might be thermal
emission associated with the accretion disk (at least in radio-quiet AGN),
then the ratio $L_{\rm r}/L_{\rm opt}$ could actually be related to the ratio
$L_{\rm non-thermal}/L_{\rm thermal}$. Furthermore, $L_{\gamma}$ seems to
scale with $L_{\rm r}$, although the details of this dependence are still
under debate (see e.g., Mattox et al.~\cite{m97} and references therein). It
could then be that even radio-quiet AGN are $\gamma$-ray emitters, although
scaled down by their ratio between radio and optical powers, that is at a
level $3 - 4$ orders of magnitude below that of blazars, with typical fluxes
$F_{\gamma} \approx 10^{-10}$ photons cm$^{-2}$ s$^{-1}$. In other words,
radio-quiet AGN would fulfill requirements number one and two \footnote{One
could argue that the relativistic beaming requirement would probably not be
very important in radio-quiet AGN as the luminosity/radius ratio in the
$\gamma$-ray band in these sources would be much lower anyway and $\gamma$-ray
photons would escape even without beaming. However, GeV photons collide
preferentially with X-ray photons, which are plentiful in these sources: some
beaming might then be required for the (putative) $\gamma$-ray emission in
radio-quiet AGN as well.} but not number three. 

Alternatively, it could be that for some reason the emission mechanisms at 
work in radio-loud sources are simply not present in the radio-quiet ones, 
either because there is no jet at all in radio-quiet AGN or because, for
example, there is no accelerating mechanism. In this case, either condition
number one or number three (or both) would be missing (number two would now be
unimportant), and no $\gamma$-ray emission would be expected. 

Unfortunately, it is not going to be possible to test these two alternatives
on the basis of $\gamma$-ray data for some time: even in the first case, in
fact, the expected $\gamma$-ray fluxes are below the sensitivity of currently
planned future $\gamma$-ray missions, like GLAST.~\cite{mo97} 

\section{AGN as TeV Sources}\label{sec:tev}

So far, only emission up to a few GeV has been considered. However, TeV
astronomy is now in full swing, as we have heard at this meeting, and it is
therefore interesting to consider the situation at these energies. 

Two, possibly three, extragalactic sources have been detected at $E > 0.3$
TeV;~\cite{w97} these are all BL Lacs. That is, even at energies above those
of EGRET (and exactly for the same reasons) the only $\gamma$-ray emitting AGN
are still blazars! The difference here is that, unlike the situation in the
EGRET band where the majority of detected blazars are flat-spectrum radio
quasars, only BL Lacs have been detected so far. Furthermore, these BL Lacs
are the three nearest confirmed BL Lacs in the recent catalogue by Padovani \&
Giommi,~\cite{pg95} namely MKN 421 (redshift $z = 0.031$), MKN 501 ($z =
0.055$) and 1ES 2344+514 ($z = 0.044$), the latter needing confirmation to be
considered a firm detection. The fact that only relatively nearby BL Lacs have
been detected is probably related to absorption of TeV photons by the infrared
background (see e.g., Biller~\cite{b97} and references therein). 

Why have no flat-spectrum radio quasars been detected at TeV energies? These
sources are typically at higher redshifts and so the effect on them of the
cosmological absorption by infrared photons is more severe. However, there are
at least four strong radio sources classified as flat-spectrum quasars at $z <
0.1$, including 3C 120 and 3C 111, the latter having been looked at by the
Whipple experiment, with negative results.~\cite{k95} (3C 111, however,
although superluminal,~\cite{vc94} is lobe-dominated [$f_{\rm core}/f_{\rm
extended} \simeq 0.2$~\cite{h95}], which suggests it is an unlikely blazar. 
Also, it has not been detected by EGRET.) 

This is certainly small number statistics and definite conclusions should only 
be drawn after a larger number of relatively local flat-spectrum radio quasars
have been observed at TeV energies. However, the non-detection of
flat-spectrum radio quasars could simply mean that {\it internal} absorption
is significant in these  sources. In fact, the cross-section for photon-photon
interaction for $\sim 1$ TeV photons is maximum at $\sim 10^{14}$ Hz or $\sim
2.5 \mu$ and quasars have a larger photon density than BL Lacs at these
frequencies, be it emission from the obscuring torus \footnote{It is not clear
if the presence of an obscuring torus is required in BL Lacs as well as in
radio quasars: see discussion in Urry \& Padovani~\cite{up95} and
Padovani~\cite{pa97} and references therein.} or even the accretion disk (see
e.g., Protheroe \& Biermann~\cite{pb97}). 

The new, more sensitive projects which are underway in the field of TeV
astronomy~\cite{pu97} will certainly shed light on these issues.
Furthermore, the detection of (necessarily local) AGN at $E > 10$ TeV would
essentially rule out leptonic models (i.e., inverse Compton emission by 
electrons) and turn the balance in favour of hadronic processes being
responsible for the $\gamma$-ray emission in these sources.~\cite{ma97} 
These so-called ``proton blazars'' would also be strong neutrino
sources~\cite{z97} possibly detectable by the large, high-energy neutrino
experiments discussed at this meeting.~\cite{mos97} 

\section{Summary}\label{sec:sum}
The main conclusions are as follows: 

\begin{enumerate}
\item The only AGN detected at GeV (EGRET) and even TeV (e.g., Whipple) energies
are blazars, that is a special class of sources which includes BL Lacertae
objects and radio quasars with a relatively flat radio spectrum. These sources
are thought to have their jets moving at relativistic speeds almost directly
towards the observer, a phenomenon which goes under the name of ``relativistic
beaming'' and causes strong amplification and collimation of the radiation in
our rest frame. 
\item As blazars make up {\it at most} 5\% of all AGN, they must have some 
peculiar characteristics which favour their $\gamma$-ray detection. I have shown
that relativistic beaming plus a strong non-thermal (jet) component play, in
fact, a fundamental role in making these sources detectable at $\gamma$-ray
energies. 
\item $\gamma$-ray missions $\sim 1,000$ times more sensitive than EGRET (that
is, below the sensitivity of presently planned missions like GLAST) might
also detect the bulk of the more common radio-quiet AGN, under the assumption
that they also possess, on much smaller scales, a non-thermal engine. 
\item TeV astronomy, a very young branch of astronomy which has already
produced some very exciting results, will likely play an important role in the
near future in constraining blazar models. 
\end{enumerate}

In summary, there exists a tight connection between unified schemes and 
$\gamma$-ray emission in AGN, as they both depend on relativistic beaming, the
former as a mechanism to produce a strong angle dependence of the observed
properties, the latter as a way to let $\gamma$-ray photons escape from the
source. 

\section*{Acknowledgments}
I acknowledge various participants in the Moriond meeting as well as Gabriele 
Ghisellini for useful comments, and Annalisa Celotti and Christina Pomel for a
careful reading of the manuscript. My participation in the Moriond meeting was
supported by ASI. 


\end{document}